\documentclass[10pt,final,journal,letter,oneside,twocolumn]{IEEEtran}
\usepackage{cite}
\usepackage{ulem}
\usepackage{soul,color}
\usepackage{srcltx}
\usepackage{multirow}
\usepackage[tbtags,sumlimits,nointlimits,reqno]{amsmath}
\usepackage{graphicx,dblfloatfix}
\usepackage{float}
\usepackage{subfigure}
\usepackage{psfrag}
\usepackage{color}
\usepackage{amssymb}
\usepackage{makecell}
\usepackage{array}
\usepackage{epstopdf}
\usepackage{mathtools}
\usepackage{multirow}
\usepackage{threeparttable}

\begin{document}

\title{Synthesis of Negative Group Delay \\Using Lossy Coupling Matrix}

\author {Ranjan~Das,~\IEEEmembership{Student Member,~IEEE,}
         Qingfeng~Zhang,~\IEEEmembership{Senior Member,~IEEE,}
         Jayanta~Mukherjee,~\IEEEmembership{Senior Member,~IEEE}

\thanks{Manuscript received Feb. 2017, revised May 2017}

\thanks{Q. Zhang is with the Department of Electronics and Electrical Engineering, Southern University of Science and Technology, Shenzhen, Guangdong, China 518055. (email: zhang.qf@sustc.edu.cn).}

\thanks{R. Das and J. Mukherjee are with Electrical Engineering Department, Indian Institute of Technology Bombay, India. R. Das is also with the Department of Electronics and Electrical Engineering, Southern University of Science and Technology, Shenzhen, Guangdong, China 518055.}

}

\maketitle

\begin{abstract}
In this paper, a systematic synthesis approach is proposed for achieving negative group delay responses using lossy coupling matrix. It is mathematically proved that, for a passive and reciprocal network, loss is the necessary condition to realize a negative group delay. Also, the optimum strategy is to place zeros and poles of the transfer function both on the left complex plane. A closed-form relation between the group delay and magnitude is then derived based on this strategy, and followed by a complete synthesis approach using coupling matrix. Two numerical and one experimental examples are finally given to illustrate the proposed synthesis method.
\end{abstract}
\IEEEoverridecommandlockouts

\IEEEpeerreviewmaketitle

\section{Introduction}

 The negative group delay (NGD) phenomenon was initially investigated by Brillouin~\cite{p7} in the 1960s, with the electromagnetic (EM) wave propagation through absorptive and dispersive media. The NGD refers to the phenomenon whereby an electromagnetic wave traverses a dispersive material or electronic circuit in such a manner that its amplitude envelope is advanced through media rather than undergoing delay. This phenomenon occurs when the linear circuit or media exhibit a positive phase slope for a specific narrow frequency band. However, NGD phenomenon does not violate causality as the front velocity never exceeds the speed of light~\cite{p7}.  In early 1990s, researchers verified that certain class of passive circuits can exhibit NGD behaviour~\cite{p1,p2,p3,p4,p5,p6}. Recently, similar NGD effect was observed with metamaterial-based passive circuits showing bandstop response wherein the group delays can be negative~\cite{p4,p5,p6}.

 Earlier NGD applications were limited to feedback element in low-frequency operational amplifier because of its high losses~\cite{p8,p9}. Recently, interesting characteristics of NGD networks have been applied to various practical applications in communication systems such as shortening or reducing transmission delays in high-speed interconnects~\cite{p10,p11}. Modern RF wireless communication systems require highly linear high-power power amplifiers (PA) because of the complex modulation techniques that are needed to handle the higher data rate transmissions. A pre-distortion is one of the popular linearization techniques~\cite{p12}, with which it is crucial to match group delays (GDs) of different paths of the pre-distortion circuit. NGD element would be a potential solution to compensate the positive group delays for improved bandwidth. In~\cite{p13,p14}, it was reported that feed-forward linear amplifier accompanying with NGD elements can improve the efficiency of the amplifier. NGD element is also employed in phased array antennas to prevent beam squinting~\cite{p15,p16}. It was shown in~\cite{p16} that, a beam-squint free operation of the series-fed antenna arrays can be achieved by using NGD elements. In~\cite{p17}, NGD network is used to realize non-Foster reactive elements, which are further applied to varactor diodes for an increased tuning range~\cite{p18}. In~\cite{p19}, an artificial magnetic conductor loaded with NGD networks significantly improves the bandwidth.

Realizations of NGD can be classified either by passive lossy circuit and active circuit, or by reflection type and transmission type. One of the early design that exhibiting NGD response operates in reflection mode, using a Lange coupler and FET-based resonators~\cite{p1}. This monolithic MIC based design suffers from performance degradation due to process variations. In~\cite{p1} a single-stage reflection-type MESFET-based tunable NGD implementation has been proposed. High negative group delay (NGD) values are obtained at 1 GHz at the expense of high signal attenuation. Later, researchers proposed distributed transmission line based NGD structures using passive lossy resonators~\cite{p25,p26,p3}, to avoid the complicated monolithic process. These NGD circuits are typically achieved by transmission-line loaded RLC resonators~\cite{p20,p21}, or bandstop filter architectures~\cite{p23}, or hybrid coupler integrated lossy resonators for improved attenuations~\cite{p22}. Most of the passive NGD circuits exhibit a large attenuation~\cite{p30}, which may be compensated by cascading unilateral~\cite{p24} or bilateral~\cite{p27} active elements. Multistage NGD structure is another popular solution to minimize the signal attenuation. An analysis of delay-bandwidth-product limit has been provided in~\cite{p28} for the design of multi-stage NGD circuits. NGD can also be realized by lossy metamaterials~\cite{p29} and lossy filters~\cite{p23,p32}.

In this paper, we propose a systematic coupling matrix synthesis approach for NGD circuits. Main contributions of this work include: 1) lossy coupling matrix synthesis for NGD is presented for the first time; 2) necessary condition and optimum strategy for achieving NGD is mathematically proved; 3) closed-form relation between the group delay and magnitude is derived; 4) condition for minimum-order network realization is presented for the first time; 5) tradeoff between transmission and reflection magnitudes are discussed under the minimum-order condition; 6) non-minimum-order network realization is also provided and discussed.

This paper is organized as follows. Sec. II derives the necessary condition for NGD and optimum in lossy networks. Group delay and magnitude relation is also described in this section. Sec. III provides complete synthesis approach for NGD network using lossy coupling matrix. Two illustrative examples are subsequently provided in Sec. IV and an experimental validation is given in Sec. V. Further discussion on non-minimum-order network is provided in Sec. VI, which is followed by the conclusion in Sec. VII.

\section{Negative Group Delay}

\subsection{Necessary Condition for Passive and Reciprocal Network}

For a passive and reciprocal network, one may start with the lossless case, which satisfies the unitary condition
 \begin{subequations}\label{eq:unitary}
 \begin{align}
&S_{11}S_{11}^*+S_{21}S_{21}^*=1,~\label{eq:unitary_1}\\
&S_{22}S_{22}^*+S_{12}S_{12}^*=1,~\label{eq:unitary_2}\\
&S_{11}S_{21}^*+S_{12}S_{22}^*=0.~\label{eq:unitary_3}
 \end{align}
 \end{subequations}
The reciprocity indicates $S_{21}=S_{12}$, which, after substitution into~\eqref{eq:unitary}, leads to
 \begin{subequations}\label{eq:tem_unitary}
 \begin{align}
|S_{11}|=|S_{22}|, ~\label{eq:s11_s22}\\
\dfrac{S_{11}}{S_{22}^*}=-\dfrac{S_{21}}{S_{21}^*}. ~\label{eq:s12_s21}
 \end{align}
  \end{subequations}
\noindent Note from \eqref{eq:s11_s22} that the reflection magnitudes, $|S_{11}|$ and $|S_{22}|$, are equal under the lossless condition.

In filter design and coupling matrix synthesis~\cite{cameron2007microwave}, the scattering matrix is usually expressed by rational polynomials, i.e.
  \begin{align}\label{eq:scattering_matrix}
\left[ {\begin{array}{*{20}{c}}
{{S_{11}}}&{{S_{12}}}\\
{{S_{21}}}&{{S_{22}}} \\
\end{array}} \right]= \frac{1}{{E\left( s \right)}}\left[ {\begin{array}{*{20}{c}}
{F_{11}\left( s \right)}&{P\left( s \right)}\\
{P\left( s \right)}&{{F_{22}}\left( s \right)}
\end{array}} \right],
  \end{align}
\noindent where $s$ is the complex frequency, and $E(s)$ is a Hurwitz polynomial whose roots are distributed on the left part of $s-$plane. By applying the above polynomial expression, \eqref{eq:tem_unitary} becomes
 \begin{subequations}
 \begin{align}
&|F_{11}(s)|=|F_{22}(s)|,~\label{eq:F11_F22}\\
&\dfrac{F_{11}(s)}{F_{22}(s)^*}=-\dfrac{P(s)}{P(s)^*}.~\label{eq:P}
 \end{align}
 \end{subequations}
The magnitude equality in \eqref{eq:F11_F22} indicates that the roots of $F_{11}(s)$ and $F_{22}(s)$ are paraconjugate pairs distributed symmetrically about the imaginary axis. Thus, the sum of their phases at $s=j\omega$ should be multiple of $\pi$, i.e.
 \begin{align}
\angle F_{11}(j\omega)+\angle F_{22}(j\omega)=N\pi, ~\label{eq:f11_f22_phase}
 \end{align}
where $N$ is the order of $F_{11}(s)$ and $F_{22}(s)$. Equation \eqref{eq:P} indicates $\angle F_{11}+\angle F_{22}=\pi+2\angle P$, which, after substitution of \eqref{eq:f11_f22_phase}, becomes $\angle P(j\omega)=\frac{\pi}{2}(N-1)$. Thus, $P(j\omega)$ has a constant phase that is $\omega$-independent. Therefore, the group delay of $S_{21}(j\omega)=P(j\omega)/E(j\omega)$ is contributed only by $1/E(j\omega)$, since the derivative of $\angle P(j\omega)$ is zero.

Assume that the roots of $E(s)$ are $a_n+jb_n$ where $n=1,2,\ldots,N$. Then, the group delay of $S_{21}(j\omega)$, $\tau(\omega)$, is computed as
 \begin{align}
\tau(\omega) =\dfrac{d\angle E(j\omega)}{d\omega}= -\sum_{n=1}^{N} {\frac{a _{n}}{{a ^2_n + {{\left( {\omega  - {b_{n}}} \right)}^2}}}}. ~\label{eq:delay_lossless}
 \end{align}
Since $E(s)$ is a Hurwitz polynomial, the real part of its roots is less than zero, i.e. $a_n< 0$. Thus, each part of the sum in~\eqref{eq:delay_lossless}, $-a_n/[a_n^2+(\omega-b_n)^2]$, is always positive, leading to a positive sum, i.e. $\tau\geq 0$. In summary, lossless network always has a positive group delay. In other words, for passive and reciprocal network loss is a necessary condition for NGD.

\subsection{Optimum Strategy in Lossy Network}

Since loss is a necessary condition for NGD, the unitary condition~\eqref{eq:unitary} is not valid any more in such cases. Thus, $P(s)$ is not restricted to have paraconjugate-pair roots and hence able to contribute a nonzero group delay response. One may look into an optimum strategy for achieving NGD in such lossy cases.

Since $S_{21}$ equals to $P(s)$ divided by $E(s)$, its group delay is computed by
\begin{equation}
\tau (\omega) =  - \frac{{\partial\angle {S_{21}(j\omega)}}}{{\partial\omega }} ={\frac{{\partial\angle E(j\omega)}}{{\partial\omega }} - \frac{{\partial\angle P(j\omega)}}{{\partial\omega }}}. ~\label{eq:tao_lossy}
\end{equation}
Assume $S_{21}(s)$ is
\begin{equation}
S_{21}(s) = K_0\dfrac{\prod_{m=1}^{M}s-(c_m+jd_m)}{\prod_{n=1}^{N}s-(a_n+jb_n)}, ~\label{eq:s21}
\end{equation}
where $K_0$ is a scaling constant, $a_n+jb_n$ and $c_m+jd_m$ are the roots of $E(s)$ and $P(s)$, respectively, $N$ and $M$ are the order of $E(s)$ and $P(s)$, respectively. Substitution of \eqref{eq:s21} into \eqref{eq:tao_lossy} leads to
 \begin{align}
\tau (\omega)= -\sum_{n=1}^{N} {\frac{a _{n}}{{a ^2_n + {{\left( {\omega  - {b_{n}}} \right)}^2}}}} + \sum_{m=1}^{M} {\frac{c _{m}}{{c ^2_m + {{\left( {\omega  - {d_{m}}} \right)}^2}}}}. ~\label{eq:delay_lossy}
 \end{align}

Note that, on the right hand of~\eqref{eq:delay_lossy}, although the first part is positive due to $a_n<0$ in the case of a Hurwitz polynomial, the second part can be negative if $c_m<0$. If the contribution from the second part is larger than that from the first part, the total delay would be negative. Therefore, the optimum strategy to achieve minimum negative NGD value is to place all the roots of $P(s)$ on the left complex plane. This can be easily proved by a counterexample. Assuming one root is on the right plane, one then replaces it using its paraconjugate counterpart, which leads to a decreased group delay and meanwhile an unaffected magnitude. In summary, in the optimum strategy, both $P(s)$ and $E(s)$ are Hurwitz polynomials.

One also should note that there is a tradeoff between the group delay and magnitude when manipulating the roots $P(s)$. To better see this, one may consider the group delay at $\omega=0$ for simplicity. According to \eqref{eq:delay_lossy}, the group delay at $\omega=0$ contributed by each root of $P(s)$ is the real part divided by magnitude square. Thus, the closer to the origin the roots of $P(s)$ are placed, the more negative group delay it achieves. However, the magnitude of $S_{21}(s)$ would get small if the roots of $P(s)$ are close to the origin. It turns out that a tradeoff exists between the magnitude and group delay. Rigorous relation between them is derived in the forthcoming section.

\subsection{Relation Between Group Delay and Magnitude}

 In the optimum strategy, zeros and poles are both on the left complex plane. Thus, $\ln S_{21}(s)$ are analytic on the right part of $s$-plane, and hence its real part and imaginary part, $\ln |S_{21}(\omega)|$ and $\angle S_{21}(\omega)$, are Hilbert pairs, i.e.~\cite{hilbert}
\begin{subequations}
\begin{align}
\ln|S_{21}(\omega)|=-\dfrac{1}{\pi}\int_{-\infty}^{\infty}\dfrac{\angle S_{21}(\omega')d\omega'}{\omega-\omega'}~\label{eq:hilbert_1}\\
\angle S_{21}(\omega)=\dfrac{1}{\pi}\int_{-\infty}^{\infty}\dfrac{\ln|S_{21}(\omega')|d\omega'}{\omega-\omega'}~\label{eq:hilbert_2}
\end{align}
\end{subequations}

Since we are working on the group delay, the derivative of phase, one may consider a new function $d\ln S_{21}(s)/ds$, which is also analytic on the right part of $s$-plane. Thus, its real and imaginary parts, $d\ln |S_{21}(\omega)|/d\omega$ and $d\angle S_{21}(\omega)/d\omega=-\tau(\omega)$, are also Hilbert pairs. Their relation is formulated by replacing $\ln|S_{21}(\omega)|$ and $\angle S_{21}(\omega)$ of \eqref{eq:hilbert_1} with $\dfrac{d\ln|S_{21}(\omega)|}{d\omega}$ and $-\tau(\omega)$, respectively, leading to
\begin{align}
\dfrac{d\ln|S_{21}(\omega)|}{d\omega}&=\dfrac{1}{\pi}\int_{-\infty}^{\infty}\dfrac{\tau(\omega')d\omega'}{\omega-\omega'}\nonumber\\
&=\dfrac{1}{\pi}\left[\int_{0}^{\infty}\dfrac{\tau(\omega')d\omega'}{\omega-\omega'}+\int_{-\infty}^{0}\dfrac{\tau(\omega')d\omega'}{\omega-\omega'}\right]\nonumber\\
&=\dfrac{1}{\pi}\left[\int_{0}^{\infty}\dfrac{\tau(\omega')d\omega'}{\omega-\omega'}+\int_{0}^{\infty}\dfrac{\tau(-\omega')d\omega'}{\omega+\omega'}\right]\nonumber\\
&=\dfrac{1}{\pi}\int_{0}^{\infty}\left[\dfrac{\tau(\omega')}{\omega-\omega'}+\dfrac{\tau(-\omega')}{\omega+\omega'}\right]d\omega'~\label{eq:tau_separate}
\end{align}
The causality indicates that the phase $\angle S_{21}(\omega)$ is an odd function, and hence its derivative, $-\tau(\omega)$, is an even function, i.e. $\tau(-\omega)=\tau(\omega)$. Upon substitution of $\tau(-\omega)=\tau(\omega)$, \eqref{eq:tau_separate} is simplified to be
\begin{align}
\dfrac{d\ln|S_{21}(\omega)|}{d\omega}=
\dfrac{2\omega}{\pi}\int_{0}^{\infty}\dfrac{\tau(\omega')d\omega'}{\omega^2-\omega'^2}.~\label{eq:tau_simplify}
\end{align}

It is interesting to note from \eqref{eq:tau_simplify} that only the derivative of the magnitude logarithm links to the group delay. Thus, adding an arbitrary constant to $\ln|S_{21}(\omega)|$, equivalent to multiplying $|S_{21}(\omega)|$ by a scaling factor, does not change the relation in \eqref{eq:tau_simplify}. Accordingly, the group delay limits the magnitude shape rather than the absolute value. This can also be deduced from \eqref{eq:s21} and \eqref{eq:delay_lossy}, where $K_0$ does not contribute to the group delay.

In a similar way, \eqref{eq:hilbert_2} is reformulated as
\begin{align}
\angle S_{21}(\omega)=
\dfrac{2\omega}{\pi}\int_{0}^{\infty}\dfrac{\ln|S_{21}(\omega')|d\omega'}{\omega^2-\omega'^2}~\label{eq:phase_simplify}
\end{align}
Then, by applying $-d/d\omega$ on both sides of \eqref{eq:phase_simplify}, one obtains the group delay
\begin{align}
\tau(\omega)=
\dfrac{2}{\pi}\int_{0}^{\infty}\dfrac{\omega^2+\omega'^2}{(\omega^2-\omega'^2)^2}\ln|S_{21}(\omega')|d\omega'~\label{eq:mag_simplify}
\end{align}

The relation between group delay $\tau(\omega)$ and magnitude $|S_{21}(\omega)|$ is completely described by \eqref{eq:tau_simplify} and \eqref{eq:mag_simplify}. If one is known, the other can be computed.

\section{Coupling Matrix Synthesis}

\subsection{Condition for Minimum-Order Network}

In the synthesis of coupling matrix, the scattering parameters should be first converted into admittance parameters. The order of the finally resultant network equals to the denominator order of the admittance parameters. Therefore, one needs to minimize this denominator order for a minimum-order network.

The conversion between scattering parameters and admittance parameters is formulated by~\cite{pozar2009microwave}
  \begin{subequations}\label{eq_y21_22_n}
  \begin{align}
{Y_{21}} = \frac{{ - 2{S_{21}}}}{{(1 + {S_{11}})(1 + {S_{22}}) - {S_{12}}{S_{21}}}},\label{eq_y21_n}\\
{Y_{22}} = \frac{{(1 + {S_{11}})(1 - {S_{22}}) + {S_{12}}{S_{21}}}}{{(1 + {S_{11}})(1 + {S_{22}}) - {S_{12}}{S_{21}}}}.\label{eq_y22_n}
\end{align}
\end{subequations}
Upon substitution of \eqref{eq:scattering_matrix} and reorganization, \eqref{eq_y21_22_n} becomes
  \begin{subequations}\label{eq_y21_22}
  \begin{align}
{Y_{21}} = \dfrac{-2P}{E+F_{11}+F_{22}+\dfrac{F_{11}F_{22}-P^2}{E}}\label{eq_y21}\\
{Y_{22}} = \dfrac{E+F_{11}-F_{22}-\dfrac{F_{11}F_{22}-P^2}{E}}{E+F_{11}+F_{22}+\dfrac{F_{11}F_{22}-P^2}{E}}\label{eq_y22}
\end{align}
\end{subequations}
Note that, if $F_{11}F_{22}-P^2$ and $E$ do not have any common factor, the denominator order of \eqref{eq_y21_22} would be twice the order of $E$ after multiplying both numerators and denominators by $E$. Thus, $F_{11}F_{22}-P^2$ and $E$ should have some common factors to decrease the denominator order. This order could be minimized if $F_{11}F_{22}-P^2$ is totally dividable by $E$. In this case, the denominator order of \eqref{eq_y21_22} equals to the order of $E$.

In this minimum-order case, one may assume
\begin{align}
{F_{11}}{F_{22}} - {P^2} = EX,\label{eq:x}
\end{align}
\noindent where $X$ is a polynomial resulted from the division of $F_{11}F_{22}-P^2$ by $E$. Then, upon substitution of \eqref{eq:x}, the conversion formula in \eqref{eq_y21_22} becomes
  \begin{subequations}\label{eq_y21_22_x}
\begin{align}
{Y_{21}} &= \frac{{ - 2P}}{{E + {F_{11}} + {F_{22}} + X}}\\
{Y_{22}} &= \frac{{E + {F_{11}} - {F_{22}} - X}}{{E + {F_{11}} + {F_{22}} + X}}
\end{align}
\end{subequations}
One should note that \eqref{eq:x} is resulted from the minimum-order requirement. If higher-order network is affordable, it is unnecessary to follow this limitation. This non-minimum order case will be further discussed in Sec. VI.

In addition to the minimum-order condition, it should also follow the passivity condition, i.e. $|S_{11}|^2+|S_{21}|^2\leq 1$ and $|S_{22}|^2+|S_{12}|^2\leq 1$. Upon substitution of \eqref{eq:scattering_matrix}, this passivity condition becomes
  \begin{subequations}\label{eq:con_pass}
\begin{align}
|F_{11}|^2\leq|E|^2-|P|^2,\\
|F_{22}|^2\leq|E|^2-|P|^2.
\end{align}
\end{subequations}
A weak condition can be further derived by applying \eqref{eq:x} to \eqref{eq:con_pass}, leading to
\begin{align}
|P^2+EX|\leq\left||E|^2-|P|^2\right|.~\label{eq:passivity}
\end{align}
One should note that \eqref{eq:con_pass} is a both necessary and sufficient condition for passivity, whereas \eqref{eq:passivity} is a necessary but insufficient condition.

\subsection{Synthesis Procedure}

The polynomials, $E$, $P$, $F_{11}$ and $F_{22}$, are limited by the minimum-order condition in \eqref{eq:x}, and passvity condition in \eqref{eq:con_pass} and \eqref{eq:passivity}. These are the main restrictions that should be considered in the synthesis.

Firstly, one optimizes the roots of $E$ and $P$ to meet the group delay specification. Both of their roots should be on the left complex plane.
To start the optimization, one firstly specifies the desired negative group delay response in the lowpass frequency domain $\omega\in[-1,1]$. To ease the optimization in a digital computer, one may discretize the frequency band into $K$ frequency points, $\{\omega_1,\omega_2,\ldots,\omega_K\}$, and specify the the corresponding group delay targets at these frequencies, $\{\tau^*_1,\tau^*_2,\ldots,\tau^*_K\}$. Then, one obtains the following optimization problem
\begin{align}
\arg\min_{\tau}\sum_{k=1}^K\left\|\tau\left( \omega_k \right) - \tau _k^*\right\|, ~\label{eq:optm_pole_zero}
\end{align}
where $\tau(.)$ is evaluated by~\eqref{eq:delay_lossy}, and the optimization parameters are $a_n+jb_n$ ($n=1,2,\ldots,N$) and $c_m+jd_m$ ($m=1,2,\ldots,M$) in~\eqref{eq:delay_lossy}, corresponding to the poles and zeros of the transfer function, respectively. Since both poles and zeros should be placed on the left complex plane, as analyzed in Sec. II-B, one needs to constrain $a_n<0$ and $c_m<0$ in the optimization. The constrained optimization problem in \eqref{eq:optm_pole_zero} may be solved by various algorithms. Here, we employ the Pareto optimization using Matlab.

One should note that the magnitude cannot be simultaneously specified due to the relation in \eqref{eq:tau_simplify} and \eqref{eq:mag_simplify}. Actually, the magnitude shape has already been fixed once the group delay is specified. However, since the group delay is determined by zeros and poles only, the scaling factor $K_0$ of \eqref{eq:s21} can be chosen independently under the passivity condition. One will find that a large $K_0$ may violate the passivity condition, or leads to a large reflection. Therefore, the choice of $K_0$ comes with a tradeoff as well. This will be further illustrated by examples in the forthcoming section.

Once $E$ and $P$ are determined, only $F_{11}$, $F_{22}$ and $X$ are free parameters in \eqref{eq:x}. Usually, one expects small reflections in both $|S_{11}|$ and $|S_{22}|$. Therefore, one may optimize $X$ to minimize $|F_{11}F_{22}|$ of \eqref{eq:x}, i.e.
\begin{align}
\arg\min_{X}\left|P^2+EX\right|.~\label{eq:optimization}
\end{align}

Upon $X$ is obtained after the optimization in~\eqref{eq:optimization}, the multiplication of $F_{11}$ and $F_{22}$ is determined by \eqref{eq:x}. Then, roots of $F_{11}$ and $F_{22}$ can be picked from the roots of $F_{11}F_{22}$. One should note that many choices are available here. According to \eqref{eq_y21_22_x}, different root selections will lead to different Y parameters and hence different coupling matrices. One should maintain a balance when allocating roots to $F_{11}$ and $F_{22}$, so that their magnitudes are both minimized.

\begin{figure}[!t]
  \centering
  \psfrag{a}[c][c]{\footnotesize $\tau^\text{spec}(\omega)$}
  \psfrag{b}[c][c]{\footnotesize \shortstack{Zeros and\\poles of \eqref{eq:s21}}}
  \psfrag{c}[c][c]{\footnotesize \shortstack{Select\\$K_0$ of \eqref{eq:s21}}}
  \psfrag{d}[c][c]{\footnotesize \shortstack{Minimization of\\\eqref{eq:optimization} under passivity\\condition in \eqref{eq:passivity}}}
  \psfrag{e}[c][c]{\footnotesize $X$}
  \psfrag{f}[c][c]{\footnotesize $F_{11} \& F_{22}$}
  \psfrag{g}[c][c]{\footnotesize $Y_{21} \& Y_{22}$}
  \psfrag{h}[c][c]{\footnotesize \shortstack{Transversal\\network}}
  \psfrag{i}[c][c]{\footnotesize \shortstack{Other\\topology}}
  \psfrag{k}[c][c]{\footnotesize \eqref{eq:x}}
  \psfrag{j}[c][c]{\footnotesize Eq.}
  \psfrag{l}[c][c]{\footnotesize \eqref{eq_y21_22_x}}
  \psfrag{o}[c][c]{\footnotesize Pole-zero optimization cycle}
  \includegraphics[width=8.8cm]{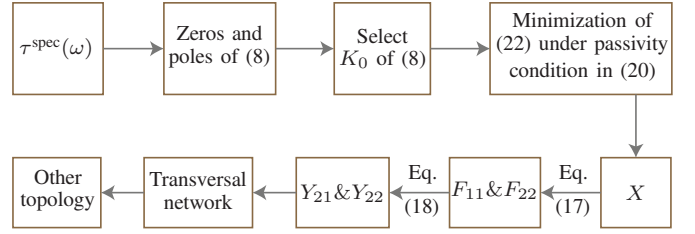}\\
  \caption{Design flow chart of the negative delay line.}\label{fig:design_steps}
\end{figure}

After the computation of $F_{11}$ and $F_{22}$, Y parameters are subsequently calculated by \eqref{eq:x}. Then, the transversal network and coupling matrix are obtained using the formulae in~\cite{mingyu_lossy,mingyu_lossy_syn}. Other topologies, e.g. folded form, are computed by applying similarity transformation provided in~\cite{cameron2007microwave}. The only difference lies in the rotation angles which are complex in lossy cases. The complete synthesis procedure is summarized in Fig.~\ref{fig:design_steps}.

\section{Illustrative Examples}

To better illustrate the synthesis procedure, let's consider two examples.

\begin{figure}[!t]
  \centering
  \psfrag{a}[c][c]{\footnotesize (a)}
  \psfrag{b}[c][c]{\footnotesize (b)}
  \psfrag{x}[c][c]{\footnotesize $\sigma$}
  \psfrag{y}[c][c]{\footnotesize $\omega$}
  \psfrag{m}[c][c]{\footnotesize $\omega$}
  \psfrag{n}[c][c]{\footnotesize $\tau(\omega)$}
  \psfrag{c}[l][c]{\footnotesize $s=\sigma+j\omega$}
  \psfrag{d}[l][c]{\footnotesize $\tau(\omega)<0$}
  \includegraphics[width=8.6cm]{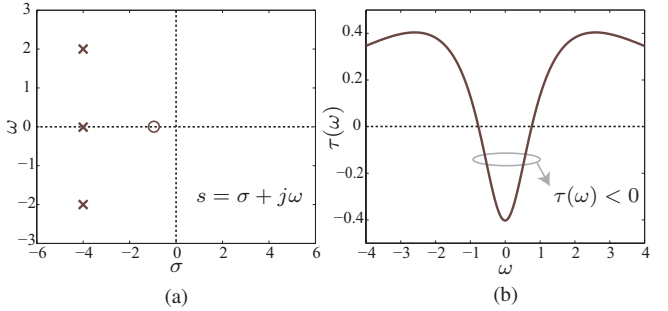}\\
  \caption{Zero-pole distribution and group delay response of \eqref{eq:s21_ex_1}.}\label{fig:s21_zp_gd}
\end{figure}

For the first example, one specifies a negative group delay $\tau=-0.4$ at $\omega=0$.
One then obtains the optimization problem in \eqref{eq:optm_pole_zero} and solves it using a Pareto optimization algorithm. Since this specification is a very weak condition, multiple solutions with different transfer function orders may exist. One tries to pick the one with a minimum order and most of the group delay below zero within $[-1,1]$. One finds that a $3^\text{rd}$-order transfer function with one zero suffices to meet the specification. The resultant $S_{21}(s)$ is
\begin{equation}
S_{21}(s) = K_0\dfrac{s+0.95}{s^3+12s^2+52s+80}, ~\label{eq:s21_ex_1}
\end{equation}
where $K_0$ is a scaling constant. Fig.~\ref{fig:s21_zp_gd} shows the corresponding pole-zero distribution and group delay response. Note that all the zeros and poles are on the left complex plane, and the resultant group delay follows the specification at $\omega=0$.
$K_0$ of \eqref{eq:s21_ex_1} should be properly chosen. Large $K_0$ increases the magnitude but leads to large reflections. To better show this, we will consider two cases, $K_0=5.6$ and $K_0=1$, and compare their performance in the following illustration.

\begin{figure}[!t]
  \centering
  \psfrag{x}[c][c]{\footnotesize $\sigma$}
  \psfrag{m}[c][c]{\footnotesize $\omega$}
  \psfrag{p}[c][c]{\footnotesize $F_{11}(s)$}
  \psfrag{q}[c][c]{\footnotesize $F_{22}(s)$}
  \includegraphics[width=6cm]{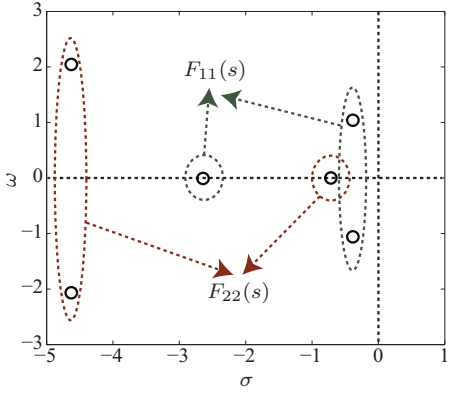}\\
  \caption{Roots distribution of $F_{11}(s)F_{22}(s)$ on the complex plane.}\label{fig:f11f22_roots}
\end{figure}

\begin{figure*}[!t]
  \centering
  \psfrag{n}[c][c]{\footnotesize $|S_{11}|$ (dB)}
  \psfrag{w}[c][c]{\footnotesize $\omega$}
  \psfrag{p}[c][c]{\footnotesize $|S_{22}|$ (dB)}
  \psfrag{m}[c][c]{\footnotesize $|S_{21}|$ (dB)}
  \psfrag{x}[l][c]{\footnotesize $K_0=5.6$}
  \psfrag{y}[l][c]{\footnotesize $K_0=1$}
  \psfrag{d}[l][c]{\footnotesize $\Delta\approx15$ dB}
  \psfrag{a}[c][c]{\footnotesize (a)}
  \psfrag{b}[c][c]{\footnotesize (b)}
  \psfrag{c}[c][c]{\footnotesize (c)}
  \includegraphics[width=17cm]{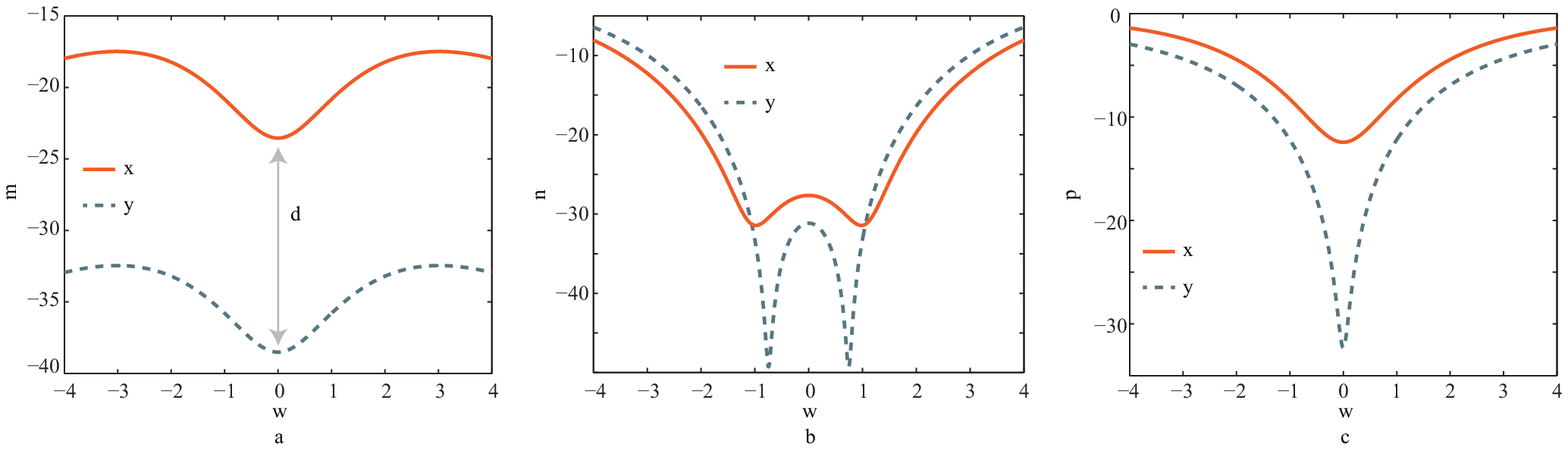}\\
  \caption{Scattering parameter comparison between $K_0=5.6$ and $K_0=1$: (a) $|S_{21}|$, (b) $|S_{11}|$, and (c) $|S_{22}|$.}\label{fig:comparison_scaling}
\end{figure*}

Firstly, one considers $K_0=5.6$. In this case, $X$ is optimized to minimize the reflections $|F_{11}F_{22}|$ under the passivity condition in \eqref{eq:passivity}. The resultant $X(s)$ is
\begin{equation}
X(s) = s^3+1.384s^2+1.098s+0.437. ~\label{eq:x_ex_1}
\end{equation}
$F_{11}(s)F_{22}(s)$ is subsequently computed by $P(s)^2+E(s)X(s)$ according to \eqref{eq:x}. Fig.~\ref{fig:f11f22_roots} shows the roots distribution of $F_{11}(s)F_{22}(s)$ on the complex plane, where three ones are close to the origin whereas other threes are far away. As we know, the closer to the origin the roots are placed, the lower reflection it achieves. To balance the roots allocation to $F_{11}(s)$ and $F_{22}(s)$, one may pick two roots close to the imaginary axis together with another one far away for $F_{11}(s)$ and the rest for $F_{22}(s)$, as shown in Fig.~\ref{fig:f11f22_roots}. The resultant $F_{11}(s)$ and $F_{22}(s)$ are
  \begin{subequations}\label{eq:f11_f22}
\begin{align}
F_{11}(s)&=s^3+3.403s^2+3.276s+3.312,\\
F_{22}(s)&=s^3+9.982s^2+32.467s+19.090.
\end{align}
\end{subequations}

As a comparison, we also compute $X(s)$, $F_{11}(s)$ and $F_{22}(s)$ in the case of $K_0=1$ using the above approach. The resultant parameters are
  \begin{subequations}\label{eq:k0=1}
\begin{align}
X(s) &= s^3-0.189s^2+0.550s-0.065,\\
F_{11}(s)&=s^3+3.872s^2+0.186s+2.216,\\
F_{22}(s)&=s^3+7.938s^2+19.353s-1.938.
\end{align}
\end{subequations}

Fig.~\ref{fig:comparison_scaling} compares the scattering parameters computed in the cases of $K_0=5.6$ and $K_0=1$, respectively. Note that, compared to $K_0=1$, $|S_{21}|$ of $K_0=5.6$ is increased by almost $15$~dB, although the reflections, $|S_{11}|$ and $|S_{22}|$, get a bit larger. It turns out that, if $K_0$ further increases, the reflection, especially $|S_{22}|$, would get worse. Therefore, there is a tradeoff between the transmission and reflection magnitudes, which is controlled by $K_0$. Here, we choose $K_0=5.6$ as our example for further illustration.

\begin{figure}[!t]
  \centering
  \psfrag{0}[c][c]{\footnotesize $0$}
  \psfrag{S}[c][c]{\footnotesize S}
  \psfrag{L}[c][c]{\footnotesize L}
  \psfrag{1}[c][c]{\footnotesize 1}
  \psfrag{2}[c][c]{\footnotesize 2}
  \psfrag{3}[c][c]{\footnotesize 3}
  \psfrag{a}[c][c][0.75]{\footnotesize $-1.446+j0.082$}
  \psfrag{b}[c][c][0.75]{\footnotesize $-1.446-j0.082$}
  \psfrag{c}[c][c][0.75]{\footnotesize $0.360$}
  \psfrag{j}[c][c][0.75]{\footnotesize $0.159-j0.366$}
  \psfrag{i}[c][c][0.75]{\footnotesize $0.159+j0.366$}
  \psfrag{d}[c][c][0.75]{\footnotesize $1.108-j0.366$}
  \psfrag{e}[c][c][0.75]{\footnotesize $-2.706-j2.342$}
  \psfrag{f}[c][c][0.75]{\footnotesize $2.706-j2.342$}
  \psfrag{g}[c][c][0.75]{\footnotesize $-j2.008$}
  \psfrag{h}[c][c][0.75]{\footnotesize $0$}
  \includegraphics[width=8.8cm]{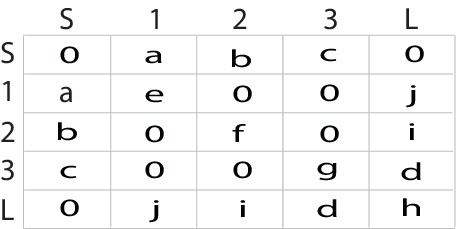}\\
  \caption{Computed transversal-form coupling matrix for the example in~\eqref{eq:s21_ex_1} and~\eqref{eq:f11_f22}.}\label{fig:transversal_ex_1}
\end{figure}

\begin{figure}[!t]
  \centering
  \psfrag{0}[c][c]{\footnotesize $0$}
  \psfrag{S}[c][c]{\footnotesize S}
  \psfrag{L}[c][c]{\footnotesize L}
  \psfrag{1}[c][c]{\footnotesize 1}
  \psfrag{2}[c][c]{\footnotesize 2}
  \psfrag{3}[c][c]{\footnotesize 3}
  \psfrag{x}[l][c]{\footnotesize Source/Load}
  \psfrag{y}[l][c]{\footnotesize Resonator}
  \psfrag{z}[l][c]{\footnotesize Real coupling}
  \psfrag{m}[l][c]{\footnotesize Lossy coupling}
  \psfrag{a}[c][c][0.75]{\footnotesize $-2.074$}
  \psfrag{b}[c][c][0.75]{\footnotesize $3.000$}
  \psfrag{c}[c][c][0.75]{\footnotesize $1.073$}
  \psfrag{d}[c][c][0.75]{\footnotesize $1.005$}
  \psfrag{h}[c][c][0.75]{\footnotesize $j1.344$}
  \psfrag{e}[c][c][0.75]{\footnotesize $-j2.034$}
  \psfrag{f}[c][c][0.75]{\footnotesize $-j3.344$}
  \psfrag{g}[c][c][0.75]{\footnotesize $-j1.314$}
  \includegraphics[width=6cm]{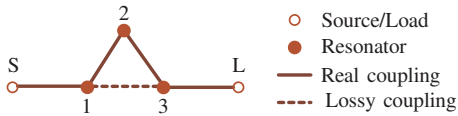}\\
  \caption{Folded-form coupling matrix and topology transformed from Fig.~\ref{fig:transversal_ex_1}.}\label{fig:fold_ex_1}
\end{figure}

\begin{figure}[!t]
  \centering
  \psfrag{x}[c][c]{\footnotesize $\tau(\omega)$}
  \psfrag{w}[c][c]{\footnotesize $\omega$}
  \psfrag{p}[c][c]{\footnotesize $|S_{22}|$ (dB)}
  \psfrag{q}[c][c]{\footnotesize $|S_{11}|$ (dB)}
  \psfrag{y}[c][c]{\footnotesize $|S_{21}|$ (dB)}
  \psfrag{a}[c][c]{\footnotesize (a)}
  \psfrag{b}[c][c]{\footnotesize (b)}
  \psfrag{c}[c][c]{\footnotesize (c)}
  \psfrag{d}[c][c]{\footnotesize (d)}
  \psfrag{n}[l][c]{\footnotesize Polynomial}
  \psfrag{m}[l][c]{\footnotesize Coupling Matrix}
  \includegraphics[width=8.8cm]{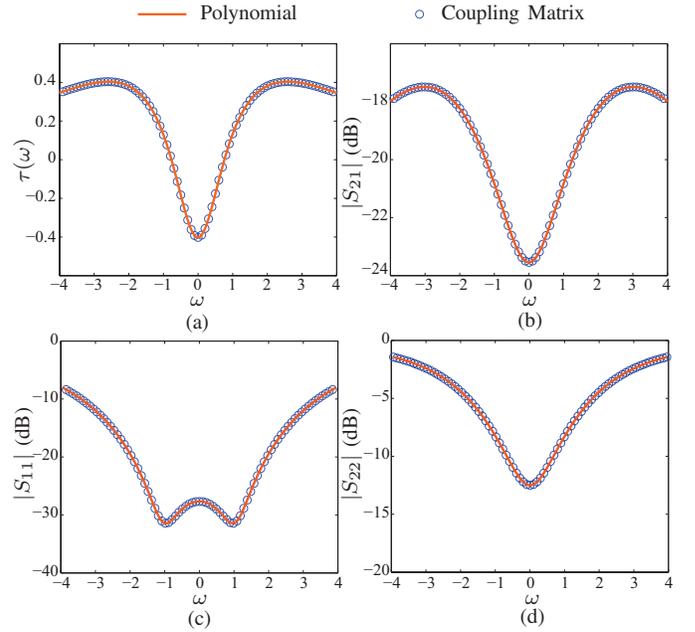}\\
  \caption{Computed responses using the coupling matrix in Fig.~\ref{fig:fold_ex_1} and polynomials in~\eqref{eq:s21_ex_1}-\eqref{eq:f11_f22}: (a) group delay, (b) $|S_{21}|$, (c) $|S_{11}|$, and (d) $|S_{22}|$.}\label{fig:response_cmx}
\end{figure}

Once $P$, $E$, $F_{11}$ and $F_{22}$ is obtained, the admittance parameters are calculated using \eqref{eq_y21_22_x}. Then, one obtains the transversal coupling matrix shown in Fig.~\ref{fig:transversal_ex_1}. It is further transformed into the folded form, as shown in Fig.~\ref{fig:fold_ex_1}, after applying the similarity transformations. One should note that the imaginary coupling coefficients represent lossy couplings. Fig.~\ref{fig:response_cmx} shows the responses computed by the coupling matrix in Fig.~\ref{fig:fold_ex_1} and polynomials in~\eqref{eq:s21_ex_1}-\eqref{eq:f11_f22}, respectively. Note that the two responses completely agrees with each other.

For the second example, one specifies a flat negative group delay $\tau=-1$ within the frequency range $[-1,1]$.
Following the same procedure, one computes the polynomials and obtains
\begin{subequations}\label{eq:poly_ex_2}
\begin{align}
P(s)&=3s^3+7.684s^2+9.539s+4.709,\\
E(s)&=s^4+15.71s^3+92.51s^2+242.2s+237.8,\\
X(s)&=s^4+1.554s^3-0.166s^2+2s+2,\\
F_{11}(s)&=s^4+3.48s^3+3.11s^2-0.05s+8.19,\\
F_{22}(s)&=s^4+13.78s^3+74.66s^2+128.6s+60.79.
\end{align}
\end{subequations}

\begin{figure}[!t]
  \centering
  \psfrag{0}[c][c]{\footnotesize $0$}
  \psfrag{S}[c][c]{\footnotesize S}
  \psfrag{L}[c][c]{\footnotesize L}
  \psfrag{1}[c][c]{\footnotesize 1}
  \psfrag{2}[c][c]{\footnotesize 2}
  \psfrag{3}[c][c]{\footnotesize 3}
  \psfrag{4}[c][c]{\footnotesize 4}
  \psfrag{x}[l][c]{\footnotesize Source/Load}
  \psfrag{y}[l][c]{\footnotesize Resonator}
  \psfrag{z}[l][c]{\footnotesize Real coupling}
  \psfrag{m}[l][c]{\footnotesize Lossy coupling}
  \psfrag{a}[c][c][0.75]{\footnotesize $-0.982$}
  \psfrag{e}[c][c][0.75]{\footnotesize $-j3.153$}
  \psfrag{f}[c][c][0.75]{\footnotesize $-j2.001$}
  \psfrag{g}[c][c][0.75]{\footnotesize $-j5.931$}
  \psfrag{i}[c][c][0.75]{\footnotesize $j2.454$}
  \psfrag{b}[c][c][0.75]{\footnotesize $1.267$}
  \psfrag{j}[c][c][0.75]{\footnotesize $-j2.292$}
  \psfrag{h}[c][c][0.75]{\footnotesize $1.528$}
  \psfrag{c}[c][c][0.75]{\footnotesize $j0.388$}
  \psfrag{k}[c][c][0.75]{\footnotesize $1.610$}
  \psfrag{n}[c][c][0.75]{\footnotesize $1.944$}
  \psfrag{d}[c][c][0.75]{\footnotesize $-5.836$}
  \includegraphics[width=7cm]{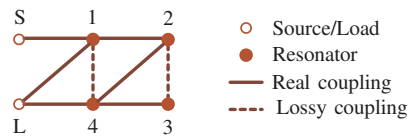}\\
  \caption{Folded-form coupling matrix and topology of the example in \eqref{eq:poly_ex_2}.}\label{fig:fold_ex_2}
\end{figure}

\begin{figure}[!t]
  \centering
  \psfrag{x}[c][c]{\footnotesize $\tau(\omega)$}
  \psfrag{w}[c][c]{\footnotesize $\omega$}
  \psfrag{p}[c][c]{\footnotesize $|S_{22}|$ (dB)}
  \psfrag{q}[c][c]{\footnotesize $|S_{11}|$ (dB)}
  \psfrag{y}[c][c]{\footnotesize $|S_{21}|$ (dB)}
  \psfrag{a}[c][c]{\footnotesize (a)}
  \psfrag{b}[c][c]{\footnotesize (b)}
  \psfrag{c}[c][c]{\footnotesize (c)}
  \psfrag{d}[c][c]{\footnotesize (d)}
  \psfrag{n}[l][c]{\footnotesize Polynomial}
  \psfrag{e}[l][c]{\footnotesize Specification}
  \psfrag{m}[l][c]{\footnotesize Coupling Matrix}
  \includegraphics[width=8.8cm]{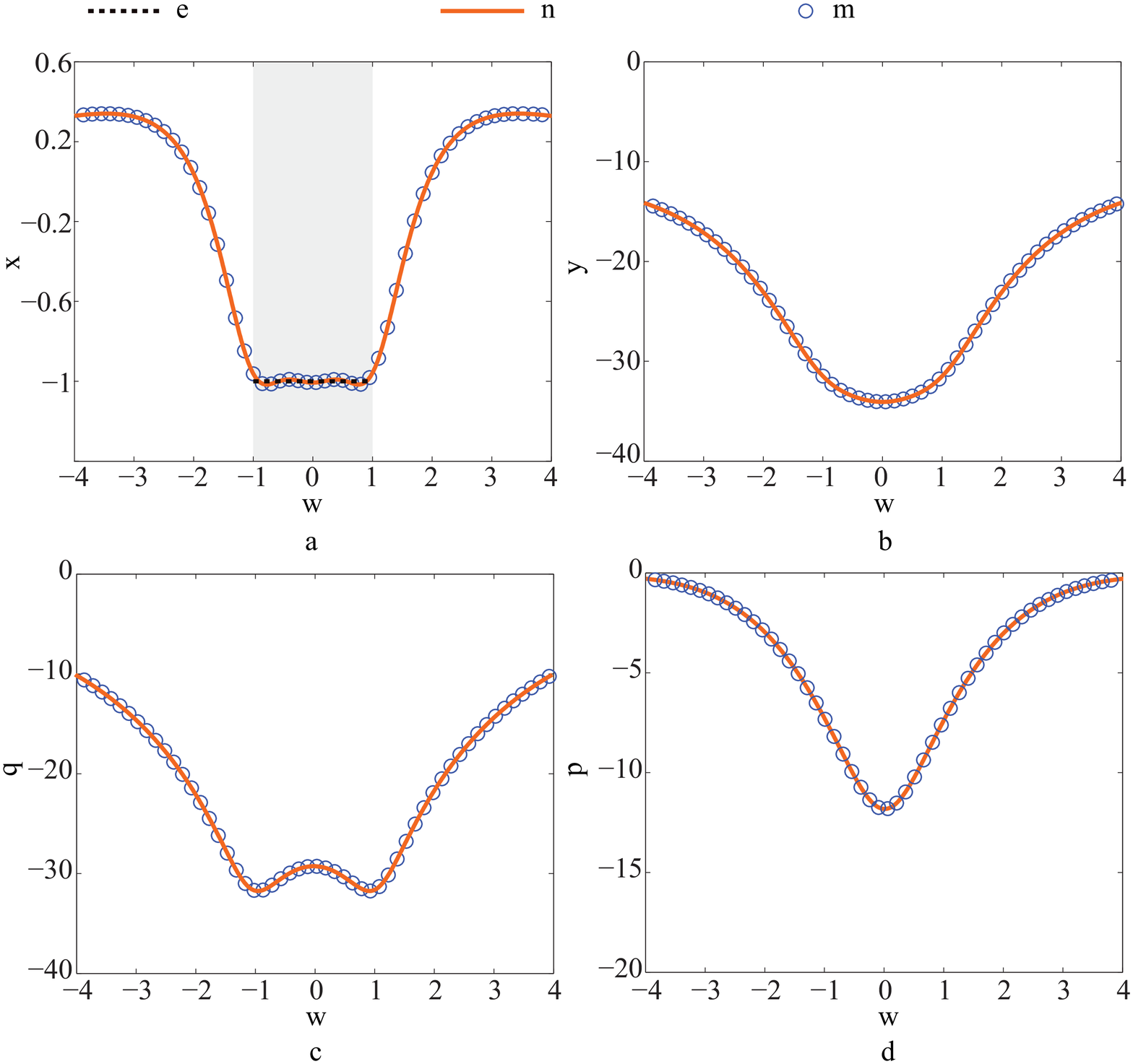}\\
  \caption{Computed responses using the coupling matrix in Fig.~\ref{fig:fold_ex_2} and polynomials in~\eqref{eq:poly_ex_2}: (a) group delay, (b) $|S_{21}|$, (c) $|S_{11}|$, and (d) $|S_{22}|$.}\label{fig:response_cmx_2}
\end{figure}

The resultant folded-form coupling matrix and topology is shown in Fig.~\ref{fig:fold_ex_2}. The computed responses are shown in Fig.~\ref{fig:response_cmx_2}. Note that the responses of the coupling matrix and polynomial completely agree with each other. The resultant group delay exhibits a negative flat response, completely following the specified one within the frequency range $[-1,1]$.

\section{Experimental Validation}

\begin{figure}[!t]
  \centering
  \psfrag{0}[c][c]{\footnotesize $0$}
  \psfrag{S}[c][c]{\footnotesize S}
  \psfrag{L}[c][c]{\footnotesize L}
  \psfrag{1}[c][c]{\footnotesize 1}
  \psfrag{2}[c][c]{\footnotesize 2}
  \psfrag{3}[c][c]{\footnotesize 3}
  \psfrag{a}[c][c][0.75]{\footnotesize $-0.4417$}
  \psfrag{b}[c][c][0.75]{\footnotesize $0.1361$}
  \psfrag{c}[c][c][0.75]{\footnotesize $0.0487$}
  \psfrag{d}[c][c][0.75]{\footnotesize $0.2140$}
  \psfrag{h}[c][c][0.75]{\footnotesize $j0.0610$}
  \psfrag{e}[c][c][0.75]{\footnotesize $-j0.0923$}
  \psfrag{f}[c][c][0.75]{\footnotesize $-j0.1518$}
  \psfrag{g}[c][c][0.75]{\footnotesize $-j0.0596$}
  \includegraphics[width=6cm]{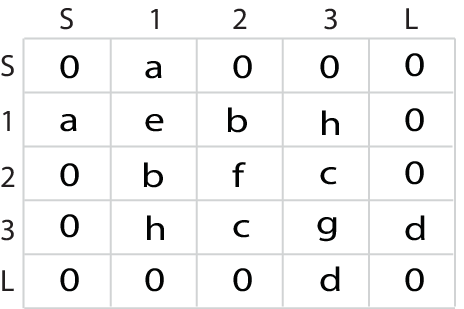}\\
  \caption{Bandpass-domain coupling matrix after scaling operation on~Fig.~\ref{fig:fold_ex_1}.}\label{fig:cmx_experiment}
\end{figure}

To further validate the proposed synthesis approach, we experimentally implement the first example in Fig.~\ref{fig:fold_ex_1}. One chooses a $50$~MHz frequency band centered at $1.73$~GHz. The synthesized coupling matrix in Fig.~\ref{fig:fold_ex_1} is then scaled according to the fractional bandwidth, leading to the bandpass-domain coupling matrix in Fig.~\ref{fig:cmx_experiment}.
Note that, both real and imaginary coupling coefficients are presented in the coupling matrix.

 We implement the resultant coupling matrix in Fig.~\ref{fig:cmx_experiment} using coupled microstrip-line resonators fabricated on an FR4 substrate ($\epsilon_r=4.3$, $\tan\sigma=0.025$) with a height of $1.6$~mm. The metallization is $20$ $\mu$m thick with a conductivity of 5.8$\times10^7$~S/m. The imaginary values in the diagonal entries [at (1,1), (2,2) and (3,3)] correspond to three resonators with finite unloaded quality factors. According to the calculation in~\cite{cameron2007microwave}, the three required unloaded quality factors are very close to $65$. For simplicity, they are all implemented using $3.1$-mm-wide half-wavelength microstrip lines, which correspond to the unloaded quality factor $65$.

\begin{figure}[!t]
  \centering
  \psfrag{u}[c][c]{\footnotesize Unit: mm}
  \psfrag{g}[c][c]{\footnotesize Gap $g$~(mm)}
  \psfrag{k}[c][c]{\footnotesize Coupling Level}
  \psfrag{a}[l][c]{\footnotesize $l_c=30$}
  \psfrag{b}[l][c]{\footnotesize $l_c=25$}
  \psfrag{c}[l][c]{\footnotesize $l_c=20$}
  \psfrag{d}[l][c]{\footnotesize $l_c=15$}
  \psfrag{e}[l][c]{\footnotesize $l_c=10$}
  \psfrag{f}[c][c]{\footnotesize $l_c$}
  \psfrag{h}[c][c]{\footnotesize $g$}
  \psfrag{x}[c][c][0.70]{\footnotesize Resonator 1}
  \psfrag{y}[c][c][0.70]{\footnotesize Resonator 2}
  \includegraphics[width=7.5cm]{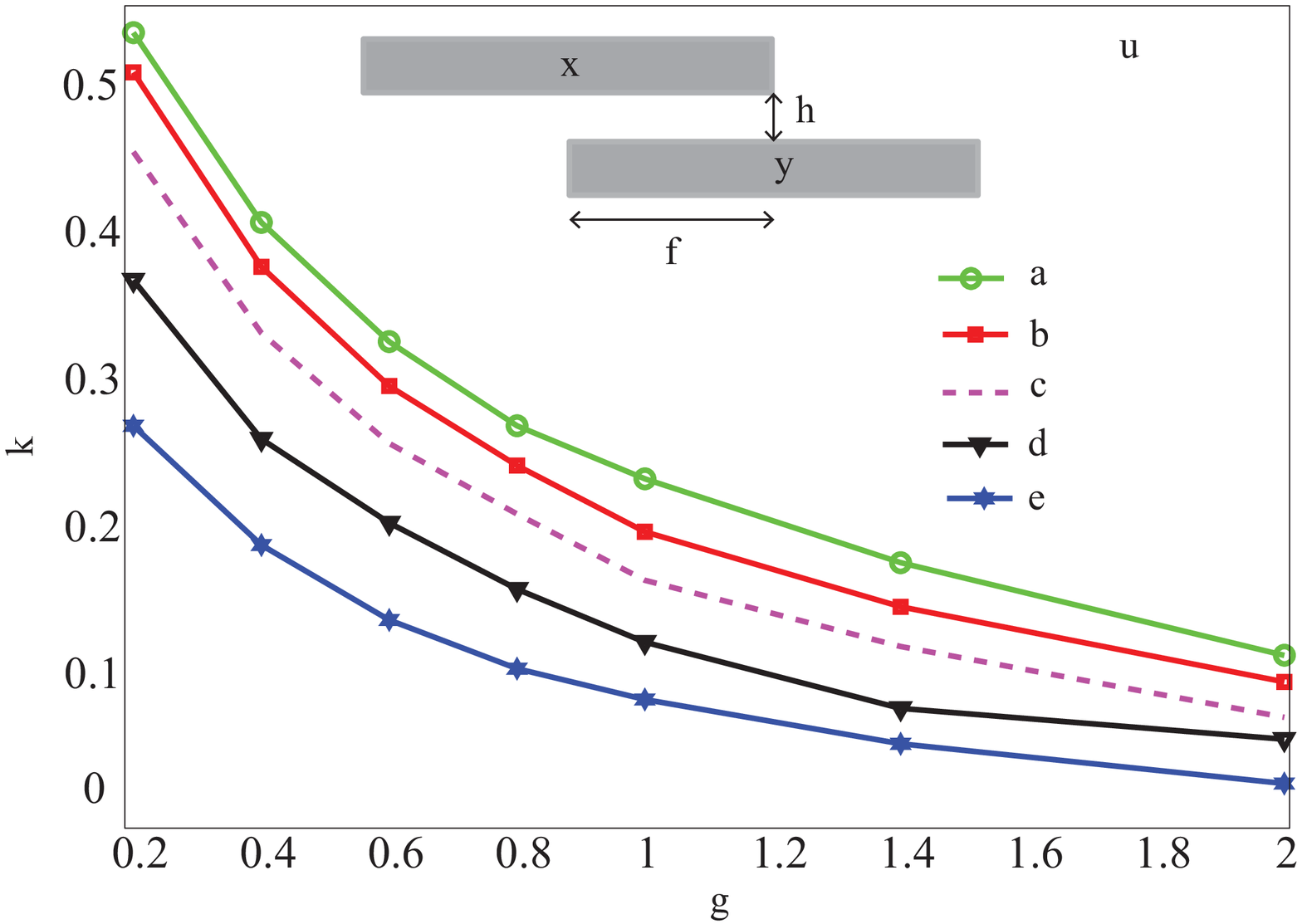}\\
  \caption{Realization of the real coupling coefficients and the design chart.}\label{fig:real_coupling}
\end{figure}

 The real coupling coefficients of Fig.~\ref{fig:cmx_experiment} [at (S,1), (1,2), (2,3) and (3,L)] are realized by gap-coupling structures, as shown in the inset figure of Fig.~\ref{fig:real_coupling}. The coupling level corresponding to such coupling structures is computed by~\cite{cameron2007microwave}
\begin{equation}
k=\dfrac{f_2^2-f_1^2}{f_2^2+f_1^2}, ~\label{eq:coupling_k}
\end{equation}
where $f_1$ and $f_2$ are two resonant frequencies of the coupled resonators, which are usually calculated using a full-wave simulation. Fig.~\ref{fig:real_coupling} plots the design chart for the coupling levels versus different coupling gap~($g$) and coupling length~($l_c$). Note that, a single coupling value has multiple solutions of $g$ and $l_c$, which provides more flexibility in the layout organization.

\begin{figure}[!t]
  \centering
  \psfrag{m}[c][c]{\footnotesize $R$}
  \psfrag{R}[c][c]{\footnotesize $\widetilde{R}$}
  \psfrag{r}[c][c]{\footnotesize $\widetilde{R}=\dfrac{R}{J^2Z_0^2}$}
  \psfrag{z}[c][c]{\footnotesize $Z_0$}
  \psfrag{j}[r][c]{\footnotesize J-Inverter}
  \psfrag{x}[c][c][0.8]{\footnotesize Resonator 1}
  \psfrag{y}[c][c][0.8]{\footnotesize Resonator 2}
  \includegraphics[width=7cm]{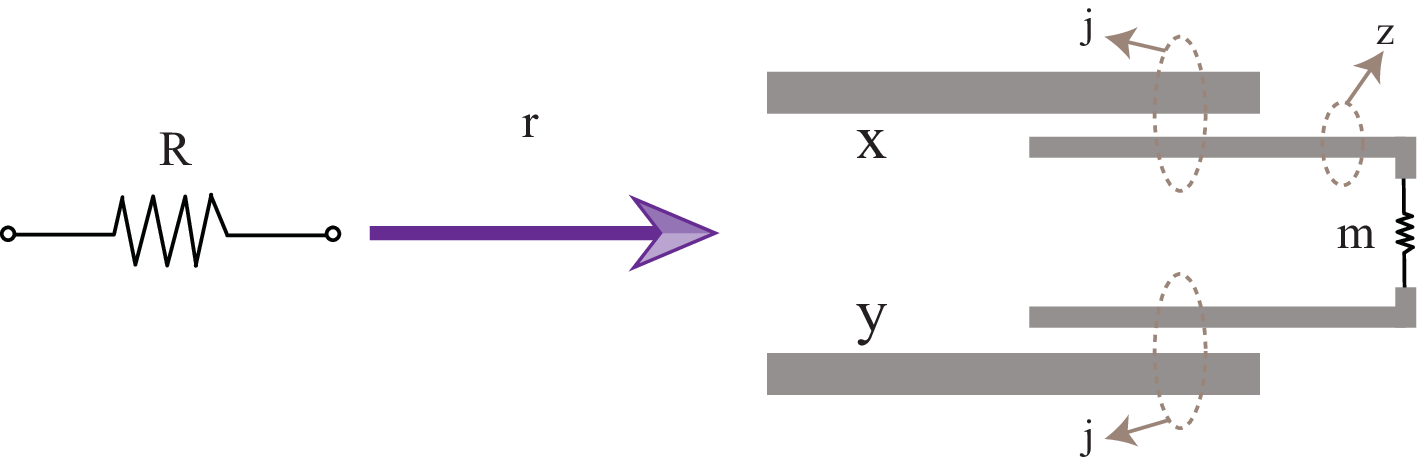}\\
  \caption{Realization of the imaginary coupling using a resistor with coupled lines.}\label{fig:R_prac}
\end{figure}

\begin{figure}[!t]
  \centering
    \psfrag{l}[c][c]{\footnotesize $l_c$~(mm)}
  \psfrag{k}[c][c]{\footnotesize Coupling Level}
  \psfrag{a}[c][c][0.75]{\footnotesize $g=0.20$,~$w=0.3$}
  \psfrag{b}[c][c][0.75]{\footnotesize $g=0.25$,~$w=0.3$}
  \psfrag{c}[c][c][0.75]{\footnotesize $g=0.20$,~$w=0.5$}
  \psfrag{d}[c][c][0.75]{\footnotesize $g=0.25$,~$w=0.5$}
  \psfrag{e}[c][c][0.75]{\footnotesize $g=0.30$,~$w=0.5$}
  \psfrag{f}[c][c]{\footnotesize $l_c$}
  \psfrag{g}[c][c]{\footnotesize $g$}
  \psfrag{h}[c][c]{\footnotesize $w$}
  \psfrag{i}[l][c]{\footnotesize $R=100~\Omega$}
  \psfrag{j}[l][c]{\footnotesize Unit:~mm}
  \includegraphics[width=7.5cm]{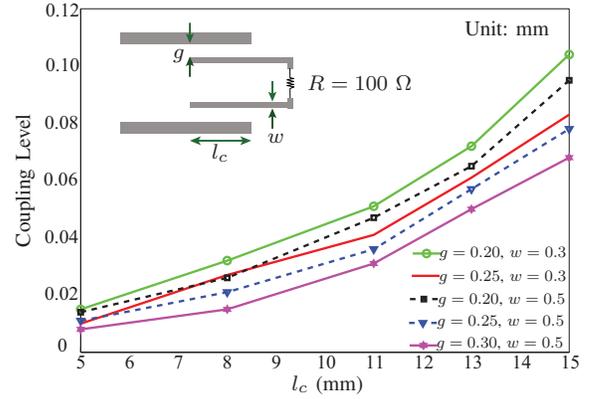}\\
  \caption{Design chart for the imaginary coupling coefficients.}\label{fig:img_coupling}
\end{figure}

The imaginary coupling coefficient of Fig.~\ref{fig:cmx_experiment} [at (1,3)] is realized using the structure in Fig.~\ref{fig:R_prac}, which is widely used in lossy filters~\cite{mingyu_lossy,mingyu_lossy_syn}. Although it could be implemented simply by a resistor, additional transmission lines and coupling structures are required in a practical layout. Two quarter-wavelength transmission lines connect the resistor on one side, and couple to two resonators through gaps (equivalent to J-inverters) on the other side. Fig.~\ref{fig:img_coupling} plots the coupling levels for different $g$, $w$ and $l_c$, which is computed using the approach in~\cite{mingyu_lossy,mingyu_lossy_syn}.

\begin{figure}[!t]
  \centering
  \psfrag{w}[][][0.75]{\footnotesize $ W=3.1$ }
  \psfrag{a}[][][0.75]{\footnotesize $ l_{p1}=16.5 $}
  \psfrag{b}[][][0.75]{\footnotesize $ g_{p1}=0.2 $}
  \psfrag{c}[][][0.75]{\footnotesize $ l_{r}=6.5 $}
  \psfrag{d}[][][0.75]{\footnotesize $ g_{r}=0.3 $}
  \psfrag{e}[][][0.75]{\footnotesize $ l_{23}=7.83 $}
  \psfrag{f}[][][0.75]{\footnotesize $ g_{23}=0.53 $}
  \psfrag{g}[][][0.75]{\footnotesize $ l_{p2}=19 $}
  \psfrag{h}[][][0.75]{\footnotesize $ g_{p2}=0.25 $}
  \psfrag{s}[l][][0.75]{\footnotesize $ s_{12}=10 $}
  \psfrag{R}[][][0.75]{\footnotesize $ R= 100 \Omega $}
  \psfrag{m}[l][][0.75]{\footnotesize $ w_{r}=0.3 $}
  \psfrag{u}[][][0.75]{\footnotesize Unit:mm}
  \includegraphics[width=88mm]{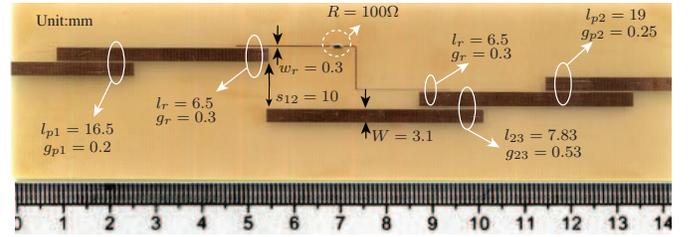}\\
  \caption{Photograph of the fabricated prototype.}\label{fig:proto_fab}
\end{figure}

\begin{figure}[!t]
  \centering
  \psfrag{x}[c][c]{\footnotesize Group Delay (ns)}
  \psfrag{w}[c][c]{\footnotesize Frequency (GHz)}
  \psfrag{p}[c][c]{\footnotesize $|S_{22}|$ (dB)}
  \psfrag{q}[c][c]{\footnotesize $|S_{11}|$ (dB)}
  \psfrag{y}[c][c]{\footnotesize $|S_{21}|$ (dB)}
  \psfrag{a}[c][c]{\footnotesize (a)}
  \psfrag{b}[c][c]{\footnotesize (b)}
  \psfrag{c}[c][c]{\footnotesize (c)}
  \psfrag{d}[c][c]{\footnotesize (d)}
  \psfrag{m}[l][c]{\footnotesize Measurement}
  \psfrag{n}[l][c]{\footnotesize Coupling Matrix}
  \includegraphics[width=8.8cm]{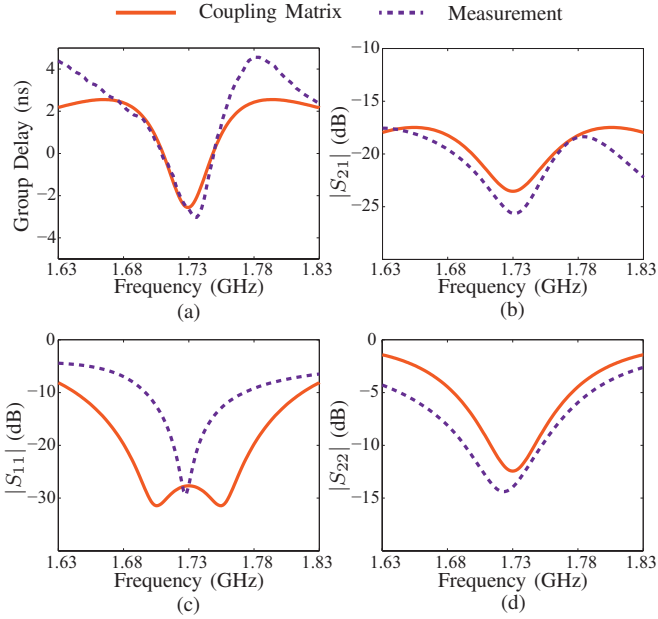}\\
  \caption{Measured responses of the fabricated prototype in Fig.~\ref{fig:proto_fab}: (a) group delay, (b) $|S_{21}|$, (c) $|S_{11}|$, and (d) $|S_{22}|$.}\label{fig:response_mea}
\end{figure}

\begin{figure}[!t]
  \centering
  \psfrag{f}[c][c]{\footnotesize Frequency (GHz)}
  \psfrag{s}[c][c]{\footnotesize $|S_{11}|$ (dB)}
  \psfrag{a}[l][c]{\footnotesize Full-wave ($g_{23}=0.65$, $l_{23}$=7.25)}
  \psfrag{b}[l][c]{\footnotesize Full-wave ($g_{23}=0.53$, $l_{23}$=7.83)}
  \psfrag{d}[l][c]{\footnotesize Measurement}
  \includegraphics[width=7cm]{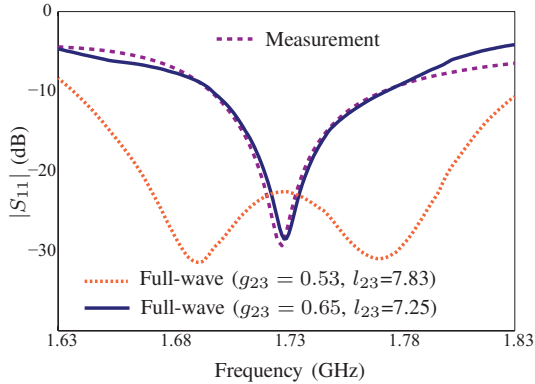}\\
  \caption{Reproduction of the measured $|S_{11}|$ in full-wave simulation by changing $l_{23}$ and $g_{23}$ (unit: mm).}\label{fig:s11_dev}
\end{figure}

Once all the initial dimensions are obtained from the design charts (Fig.~\ref{fig:real_coupling} and Fig.~\ref{fig:img_coupling}), full-wave simulations have been performed using Advanced Design System. Global full-wave optimization is finally performed to fine tune the initial parameters. Fig.~\ref{fig:proto_fab} shows the fabricated prototype with optimized dimensions. It is measured in a vector network analyzer. Fig.~\ref{fig:response_mea} shows the measured responses in comparison with the ones calculated from the coupling matrix in Fig.~\ref{fig:cmx_experiment}. Note that they generally agree with each other, except that a deviation occurs in $|S_{11}|$ response possibly due to the fabrication tolerance. In order to further investigate which physical dimensions produce this discrepancy, we carry out full-wave simulations to reproduce the measured $|S_{11}|$ response. We find that a slight change of $g_{23}$ and $l_{23}$ in Fig.~\ref{fig:proto_fab} will leads to a significant change in $|S_{11}|$ response, whereas other responses remain almost unchanged. Fig.~\ref{fig:s11_dev} shows the reproduction of the measured $|S_{11}|$ by changing $g_{23}$ and $l_{23}$ in full-wave simulation. Note that, the full-wave $|S_{11}|$ using the initial dimensions ($g_{23}=0.53$, $l_{23}$=7.83) closely follows the coupling matrix response of Fig.~\ref{fig:response_mea}(c), while the reproduced $|S_{11}|$ using the new dimensions ($g_{23}=0.65$, $l_{23}$=7.25) closely follows the measurement result. The discrepancy between the measured and full-wave responses may be further improved by a predistortion technique.

Although achieving an optimum performance is beyond the focus of this paper, we still compare our NGD circuit with previous works. Since NGD circuits always offer tradeoff among different parameters, i.e. minimum negative group delay (GD$_\text{min}$), negative group delay bandwidth~(BW) and maximum signal attenuation~(SA$_\text{max}$), we need to define a figure-of-merit (FoM) for this comparison. Here, we use the FoM definition in~\cite{p41}, i.e. FoM=$|\text{GD}_\text{min}|\times\text{BW}/\sqrt{|\text{SA}_\text{max,dB}|}$.
Tab.~\ref{comp_table} compares our work with previous works. Note that, our work is very good in terms of FoM among all the works. It could be further optimized to get a better performance.

 \begin{table}[!t]
\centering
\begin{threeparttable}[!t]
\caption{Performance comparison with previous works}
\label{comp_table}
\begin{tabular}{|c|c|c|c|c|}
\hline
Ref.  & f$_0$/BW (GHz)  & SA$_\text{max}$ (dB)   &GD$_\text{min}$ (ns) & FoM \\ \hline
\cite{p1}	  &1.00/0.013	   &32.5	    &$-10$	 	  &0.022  \\ \hline
\cite{p22}    &2.14/0.01	   &16.9	    &$-7.9$	 	  &0.019  \\ \hline
\cite{p32}    &1.00/0.1	       &38.5	    &$-1.5$	 	  &0.026  \\ \hline
\cite{p41}    &2.20/0.06	   &3.2	        &$-0.9$	 	  &0.029  \\ \hline
\cite{p33}	  &2.14/0.04	   &34	        &$-4.2$	 	  &0.028  \\ \hline
\cite{p34}	  &1.96/0.10	   &29.3	    &$-1.1$	 	  &0.021  \\ \hline
\cite{p35}	  &2.14/0.05	   &35	        &$-3.5$	 	  &0.029  \\ \hline
\cite{p36}	  &1.26/0.02	   &2.4	        &$-2.2$	 	  &0.025  \\ \hline
\cite{p39}	  &2.14/0.015	   &15.4	    &$-6.5$	 	  &0.024  \\ \hline
\cite{p40}    &2.14/0.013	   &17.5	    &$-7.5$	      &0.023  \\ \hline
\textbf{Our Work}	 &\textbf{1.73/0.05}	&\textbf{22}  &\textbf{$-3.0$}  &\textbf{0.031}  \\\hline
\end{tabular}
  \end{threeparttable}
\end{table}

\section{Discussion on Non-Minimum-Order Network}

As shown in Sec. III-A, minimum-order network follows the condition in \eqref{eq:x}. This condition limits the choices of reflection polynomials, $F_{11}$ and $F_{22}$, further leading to the magnitude tradeoff between transmission and reflection.
If minimum order is not required, all of these restrictions can be removed. It means that the transmission magnitude can be maximized under passivity condition, and the reflections can be arbitrarily suppressed. The resultant network in this case is non-minimum order.

To illustrate this, let's consider an extreme case when reflections are completely zero, i.e. $F_{11}=F_{22}=0$. In this case, the formula to compute admittance parameters in \eqref{eq_y21_22} is rewritten as
  \begin{subequations}\label{eq:discuss_conv}
  \begin{align}
{Y_{21}} = \dfrac{-2PE}{E^2-P^2},\\
{Y_{22}} = \dfrac{E^2+P^2}{E^2-P^2}.
\end{align}
\end{subequations}
  Note that the denominator order of $Y_{21}$ and $Y_{22}$ is twice that of $E$. Thus, it sacrifices the network order to achieve zero reflections.

To further illustrate this, let's revisit the example in \eqref{eq:s21_ex_1}. We will maintain the zeros and poles in \eqref{eq:s21_ex_1} to maintain the group delay response. However, since the tradeoff between transmission magnitude and reflection magnitude is removed, $K_0$ can be maximized as long as the magnitude is below $0$ dB under the passivity condition. The optimum $K_0$ is computed as $41.6$. Thus, $P(s)$ and $E(s)$ are expressed as
\begin{subequations}\label{eq:discuss_EP}
  \begin{align}
P(s) &= 41.6(s+0.95),\\
E(s) &= s^3+12s^2+52s+80.
\end{align}
\end{subequations}

\begin{figure*}[!t]
  \centering
  \psfrag{0}[c][c]{\footnotesize $0$}
  \psfrag{S}[c][c]{\footnotesize S}
  \psfrag{L}[c][c]{\footnotesize L}
  \psfrag{7}[c][c]{\footnotesize 1}
  \psfrag{8}[c][c]{\footnotesize 2}
  \psfrag{9}[c][c]{\footnotesize 3}
  \psfrag{v}[c][c]{\footnotesize 4}
  \psfrag{u}[c][c]{\footnotesize 5}
  \psfrag{z}[c][c]{\footnotesize 6}
  \psfrag{y}[c][c][0.75]{\footnotesize $0$}
  \psfrag{A}[c][c]{\footnotesize (a)}
  \psfrag{B}[c][c]{\footnotesize (b)}
  \psfrag{C}[c][c]{\footnotesize (c)}
  \psfrag{x}[c][c]{\footnotesize $\tau(\omega)$}
  \psfrag{w}[c][c]{\footnotesize $\omega$}
  \psfrag{s}[c][c][0.83]{\footnotesize $|S_{21}|$ (dB)}
  \psfrag{a}[c][c][0.75]{\footnotesize $-j1.853$}
  \psfrag{b}[c][c][0.75]{\footnotesize $1.194+j1.273$}
  \psfrag{c}[c][c][0.75]{\footnotesize $1.194-j1.273$}
  \psfrag{d}[c][c][0.75]{\footnotesize $-1.355+j0.348$}
  \psfrag{i}[c][c][0.75]{\footnotesize $-1.355-j0.348$}
  \psfrag{j}[c][c][0.75]{\footnotesize $0.622$}
  \psfrag{e}[c][c][0.75]{\footnotesize $-j11.399$}
  \psfrag{f}[c][c][0.75]{\footnotesize $-7.080-j5.229$}
  \psfrag{g}[c][c][0.75]{\footnotesize $7.080-j5.229$}
  \psfrag{h}[c][c][0.75]{\footnotesize $-1.860-j0.300$}
  \psfrag{k}[c][c][0.75]{\footnotesize $1.860-j0.300$}
  \psfrag{l}[c][c][0.75]{\footnotesize $-j1.543$}
  \psfrag{m}[c][c][0.75]{\footnotesize $j1.853$}
  \psfrag{n}[c][c][0.75]{\footnotesize $1.194+j1.273$}
  \psfrag{o}[c][c][0.75]{\footnotesize $1.194-j1.273$}
  \psfrag{p}[c][c][0.75]{\footnotesize $1.355-j0.348$}
  \psfrag{q}[c][c][0.75]{\footnotesize $1.355+j0.348$}
  \psfrag{r}[c][c][0.75]{\footnotesize $0.622$}
  \includegraphics[width=17.6cm]{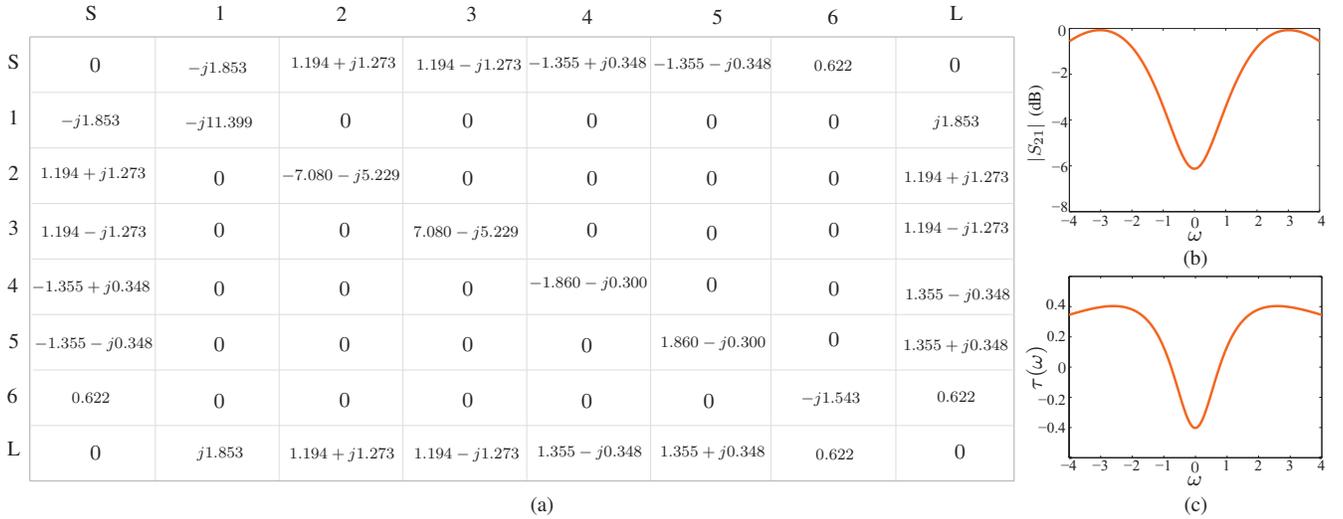}\\
  \caption{(a) Transversal-form coupling matrix, (b) magnitude and (c) group delay responses of the example in~\eqref{eq:discuss_EP}.}\label{fig:transversal_ex_discussion}
\end{figure*}

After substituting \eqref{eq:discuss_EP} into \eqref{eq:discuss_conv}, one obtains the admittance parameters and subsequently calculates the coupling matrix. Fig.~\ref{fig:transversal_ex_discussion} shows the corresponding coupling matrix and responses. Note from the coupling matrix that the network order is twice the original one of Fig.~\ref{fig:transversal_ex_1}. However, the magnitude increases about $17.5$ dB compared with the original one of Fig.~\ref{fig:response_cmx}(b), while the group delay response maintain the same as Fig.~\ref{fig:response_cmx}(a). The reflection responses are not shown here because they are completely zero.
Therefore, if an increased network order is affordable, one could significantly improve the transmission magnitude and suppress the reflections.

\section{Conclusion}
In this paper, a systematic synthesis approach has been proposed for NGD.It was mathematically proved that loss is the necessary condition, the optimum strategy was to place zeros and poles of the transfer function both on the left complex plane. A closed-form relation between the group delay and magnitude has been derived. The synthesis procedure was described. Two numerical and one experimental examples were finally given to illustrate the proposed synthesis method.

\bibliographystyle{IEEEtran}
\bibliography{mybib}

\end{document}